\documentclass[aps,physrev,groupedaddress,twocolumn,10pt]{revtex4-2}

\usepackage{amsthm,amssymb}
\usepackage{physics}
\usepackage{graphicx}
\usepackage{xcolor}
\usepackage{tikz}
\usetikzlibrary{shapes.geometric, arrows}
\usetikzlibrary{positioning}
\usepackage{subcaption}
\usepackage{siunitx}

\definecolor{purple_et}{rgb}{0.5, 0, 1}

\definecolor{blue_mod}{RGB}{4,90,122}

\begin{document}
\title{Adaptive Entanglement Witnessing for Two-Qubit Systems} 
\author{Lev Gruber}
\altaffiliation[Present address: ]{Department of Physics, Boston College, 140 Commonwealth Avenue, Chestnut Hill, MA, 02467, USA}
\affiliation{Department of Physics, Harvey Mudd College, 301 Platt Blvd., Claremont, CA 91711, USA}
\author{Ria Haapala}
\affiliation{Department of Physics, Harvey Mudd College, 301 Platt Blvd., Claremont, CA 91711, USA}
\author{Elio Thadhani}
\affiliation{Department of Physics, Harvey Mudd College, 301 Platt Blvd., Claremont, CA 91711, USA}
\author{Paco Navarro}
\affiliation{Department of Physics, Harvey Mudd College, 301 Platt Blvd., Claremont, CA 91711, USA}
\author{Eritas Yang}
\altaffiliation[Present address: ]{Department of Astrophysical Sciences, Princeton University, 4 Ivy Lane, Princeton, NJ 08540, USA}
\affiliation{Department of Physics, Harvey Mudd College, 301 Platt Blvd., Claremont, CA 91711, USA}
\author{Benjamin Hartley}
\affiliation{Department of Physics, Harvey Mudd College, 301 Platt Blvd., Claremont, CA 91711, USA}
\author{Rebecca Verghese}
\affiliation{Department of Physics, Harvey Mudd College, 301 Platt Blvd., Claremont, CA 91711, USA}
\author{Oscar Scholin}
\altaffiliation[Present address: ]{Department of Physics, University of Oxford, Oxford OX1 3PU, UK}
\affiliation{Department of Physics, Harvey Mudd College, 301 Platt Blvd., Claremont, CA 91711, USA}
\author{Isabel Godoy}
\affiliation{Department of Physics, Harvey Mudd College, 301 Platt Blvd., Claremont, CA 91711, USA}
\author{Stuart Kerr}
\affiliation{Department of Physics, Harvey Mudd College, 301 Platt Blvd., Claremont, CA 91711, USA}
\author{Brayden Mendoza}
\affiliation{Department of Physics, Harvey Mudd College, 301 Platt Blvd., Claremont, CA 91711, USA}
\author{Alec Roberson}
\affiliation{Department of Physics, Harvey Mudd College, 301 Platt Blvd., Claremont, CA 91711, USA}
\author{Theresa W. Lynn}
\email{lynn@hmc.edu}
\affiliation{Department of Physics, Harvey Mudd College, 301 Platt Blvd., Claremont, CA 91711, USA}

\date{\today}

\begin{abstract}
Entanglement witnesses are a key instrument in determining the entanglement of unknown bipartite qubit systems without requiring a full tomography. Riccardi \textit{et al.}~(2020) proposed a set of two-qubit witnesses, denoted $\{W^3\}$, which require a set of local measurements in three bases. In this work we propose an additional set of witnesses, denoted $\{W^5\}$, that build on $\{W^3\}$ and require measurements in two additional bases. We present an adaptive two-step method that uses the measurement results obtained for $\{W^3\}$ to select which $W^5$ to evaluate. We computationally demonstrate that our adaptive two-step protocol can identify $72\%$ of entangled states generated according to the methods of Roik \textit{et al.}~(2021) while requiring measurements in only five of the nine bases needed for full quantum tomography. Additionally, we experimentally verify the behavior of both sets of witnesses using polarization-entangled photon pairs.
\end{abstract}

\maketitle

\section{Introduction}
Quantum entanglement is a fundamental property of quantum mechanics and lies at the heart of multiple applications of quantum information theory. Detecting entanglement plays an important role in ensuring the security of numerous cryptographic protocols \cite{ekert1991quantum,simon2014entangled, PhysRevLett.68.557}. To do so, one could take a full tomography of a shared resource to fully recreate its density operator; in the two-qubit case, this requires correlated local measurements in nine bases. However, this approach can be inefficient and time-consuming, particularly when extended to larger systems or employed in long-distance entanglement distribution, where current realizations have low bit rates \cite{schmale2022efficient,satground-entdist-2017,entdist-qmem-2021,qkd-dlevel-timeent-2025,qkd-freqent-2025,qkd-polent-fiber-2026,freqent-fibercert-2026}. As an alternative, entanglement witnesses offer a more efficient method of detection, requiring fewer measurements than full tomography \cite{chruscinski2014entanglement}. Notably, this advantage is most significant if we hold prior knowledge about the state being evaluated, which reflects the typical use case of entanglement detection \cite{TERHAL2002313}. In contrast, here we consider situations in which there is no prior knowledge of the state.

In this paper, we consider entanglement witnesses of the form 
\begin{equation}
    W = (\dyad{\psi})^\Gamma,\label{eq:extremDecom}
\end{equation}
where $\ket{\psi} $ is a bipartite entangled state and $\Gamma$ denotes the partial transpose. In general, a two-qubit state $\rho$ is separable if and only if $\expval{W} = \Tr\{W\rho\} \geq0$ for all entanglement witnesses $W$ \cite{HORODECKI19961,chruscinski2014entanglement}. Witnesses of the form of Eq.~\ref{eq:extremDecom} witness entangled states that violate the Peres-Horodecki separability criterion which encompasses all two-qubit entangled states \cite{chruscinski2014entanglement, guhne2009entanglement, PhysRevLett.77.1413, HORODECKI19961}. These witnesses, referred to as extremal decomposable witnesses, are optimal, meaning that there exists no other witness whose set of detected states is a super-set of those detected by $W$ \cite{chruscinski2014entanglement,guhne2009entanglement}. As a result, every two-qubit entangled state is witnessed by some extremal decomposable witness, but not all entangled states are witnessed by a given witness. To detect entanglement in unknown quantum states, it is necessary to test entanglement witnesses that are capable of detecting different subsets of entangled states.

In this work, we are motivated by the goal of maximizing the probability of witnessing entanglement while minimizing the number of measurements required on many copies of an unknown bipartite state. We propose an adaptive protocol to identify different entangled states using iterative measurements to evaluate several classes of entanglement witnesses. The first step involves determining the expectation values of a set of witnesses denoted $\{W^3\}$. The $W^3$ witnesses, defined in Sec.~\ref{sec:W3}, are a set of optimal decomposable witnesses identified by \textcite{riccardi_optimal_2020}. Computing the expectation values of these witnesses requires correlated local measurements in only three bases, as opposed to the nine involved in full tomography. If any $W^3$ expectation value is negative, we have witnessed entanglement and the protocol has succeeded. If all expectation values are positive, then we proceed to the second step of our protocol.

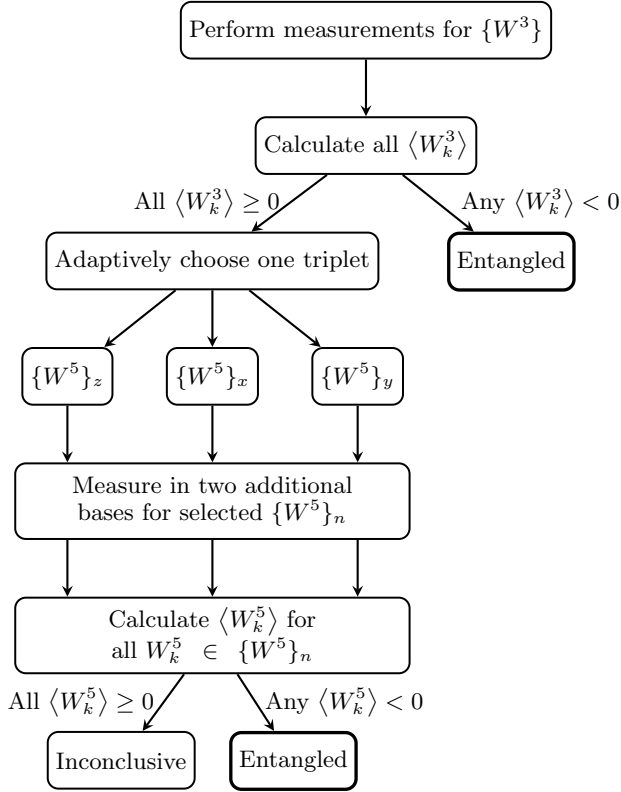
\begin{figure}
    \centering
    \tikzset{every picture/.style={line width=.75pt}} 
    \definecolor{mygreen}{RGB}{14, 102, 64} 
\tikzstyle{flow} = [rectangle, rounded corners, 
minimum width=1cm, 
minimum height=0.75cm,
text centered, 
draw=black]
\tikzstyle{shadow} = [rectangle, rounded corners, 
minimum width=1cm, 
minimum height=1cm,
text centered]
\tikzstyle{measaditional} = [rectangle, rounded corners, 
minimum width=1cm, 
minimum height=1cm,
text width=5cm,
text centered, 
draw=black]
\tikzstyle{ent} = [rectangle, rounded corners, 
minimum width=1cm, 
minimum height=0.75cm,
text centered, 
line width=1.25pt,
draw=black]
\tikzstyle{arrow} = [thick,->,>=stealth]
    \begin{tikzpicture}[node distance=0.75cm and 0.1cm, yscale=0.5, xscale=0.5]
    \node (meas) [flow] {Perform measurements for $\{W^3\}$};
    \node (w3) [flow, below=of meas] {Calculate all $\expval{W^3_k}$};
    \node (ent1) [ent, below right=of w3,  xshift=-.5cm] {Entangled};
    \node(adapt) [flow, below left=of w3, xshift=1.75cm] {Adaptively choose one triplet};
    \node (w5z) [flow, below left=of adapt, xshift=1cm] {$\{W^5\}_z$};
    \node (w5x) [flow, below =of adapt] {$\{W^5\}_x$};
    \node (w5y) [flow, below right =of adapt, , xshift=-1cm] {$\{W^5\}_y$};
    \node (smz) [shadow, below=of w5z] {};
    \node (smx) [shadow, below=of w5x] {};
    \node (smy) [shadow, below=of w5y] {};
    \node (measad) [measaditional, below =of w5x] {Measure in two additional bases for selected $\{W^5\}_n$};
    \node (scz) [shadow, below=of smz] {};
    \node (scx) [shadow, below=of smx] {};
    \node (scy) [shadow, below=of smy] {};
    \node (calc5) [measaditional, below =of measad] {Calculate $\expval{W^5_k}$ for all $W^5_k \in \{W^5\}_n$};
    \node (ent2) [ent, below right =of calc5, xshift=-2.5cm] {Entangled};
    \node (incon) [flow, below left =of calc5, xshift=+2.5cm] {Inconclusive};
    
    \draw [arrow] (meas) -- (w3);
    \draw [arrow] (w3) -- node[anchor=west] {\: Any $\expval{W^3_k}<0$} (ent1);
    \draw [arrow] (w3) -- node[anchor=east] {All $\expval{W^3_k}\geq0$} (adapt);

    \draw [arrow] (adapt) -- (w5z);
    \draw [arrow] (adapt) -- (w5x);
    \draw [arrow] (adapt) -- (w5y);

    \draw [arrow] (w5z) -- (smz);
    \draw [arrow] (w5x) -- (smx);
    \draw [arrow] (w5y) -- (smy);

    \draw [arrow] (smz) -- (scz);
    \draw [arrow] (smx) -- (scx);
    \draw [arrow] (smy) -- (scy);

    \draw [arrow] (calc5) -- node[anchor=east] {All $\expval{W_k^5}\geq0$} (incon);

    \draw [arrow] (calc5) -- node[anchor=west] {Any $\expval{W_k^5}<0$} (ent2);

    \end{tikzpicture}
  \caption{An adaptive choice protocol for entanglement witnessing with limited local measurements. The first step involves local measurements in three bases to compute all $\expval{W^3_k}$. At the adaptive choice step, the protocol chooses one subset of $\{W^5\}$ to test, which involves taking local measurements in two additional bases. When the protocol terminates, it has either certified entanglement or generated an inconclusive result. An inconclusive result does not indicate a state is not entangled; rather, it indicates the measurements performed were insufficient to certify entanglement.}
    \label{fig:flowchart}
\end{figure}

The second step of the adaptive protocol uses our additional entanglement witnesses, denoted $\{W^5\}$. These witnesses are split into three subsets and are derived by performing geometric rotations on $\{W^3\}$, as explained in detail in Sec.~\ref{sec:W5}. We adaptively choose one of the three $\{W^5\}_{n} : n \in \{z,x,y\}$ subsets using the methods described in Sec.~\ref{sec:adaptive_choice} and perform the additional associated measurements. If any of the $\{W^5\}_{n}$ witnesses have a negative expectation value, then the protocol has successfully verified entanglement using measurements in five bases. See Fig.~\ref{fig:flowchart} for a flow chart of this process.

Within the framework of this adaptive protocol, we present two methods for making the second-stage adaptive choice, analyze our protocol's success rate, and experimentally verify our protocol using pairs of polarization-entangled photons. Furthermore, we address the strengths and weaknesses of our model and present potential extensions of the adaptive witnessing scheme.

\section{$\{W^3\}$ Witnesses}\label{sec:W3}
\subsection{Construction}
The set of optimal, extremal decomposable witnesses proposed by \textcite{riccardi_optimal_2020} is defined by six families of entangled $1$-parameter, rank-$1$ projectors as
\begin{equation}
    \{W^3\} = \left\{W_k^3=\left(\dyad{\varphi^3_k}\right)^\Gamma \mid k = 1,\dots 6\right\} \label{eq:RicardiW3Set}
\end{equation}
where
\begin{equation}
\begin{aligned}
    \ket{\varphi^3_1} =& \cos \theta \ket{\Phi^+}+\sin\theta\ket{\Phi^-},\\ 
    \ket{\varphi^3_2}=& \cos \theta\ket{\Psi^+}+\sin \theta \ket{\Psi^-},\\
    \ket{\varphi^3_3} =& \cos \theta\ket{\Phi^+}+\sin\theta\ket{\Psi^+},\\
    \ket{\varphi^3_4} =& \cos \theta\ket{\Phi^-}+\sin \theta\ket{\Psi^-},\\
    \ket{\varphi^3_5} =& \cos \theta\ket{\Phi^+}+i\sin \theta\ket{\Psi^-},\\ 
    \ket{\varphi^3_6} =& \cos \theta\ket{\Phi^-}+i\sin \theta\ket{\Psi^+}.
\end{aligned}
\label{eq:w3phistates}
\end{equation}
Here $\ket{\Phi^{\pm}}$, $\ket{\Psi^{\pm}}$ are the standard two-qubit Bell states and $\theta \in [0,\pi]$. Note that valid witnesses are only obtained for $\theta$ values such that $\ket{\varphi^3_k}$ is entangled; otherwise $\Tr\{W^3\rho\}\geq 0$ for all $\rho$.

These witnesses require correlated local measurements in three bases to determine the Stokes parameters $\{S_{ii}=\expval{\sigma_i \otimes \sigma_i}, S_{0i}=\expval{\mathbb{I}\otimes \sigma_i}, S_{i0}=\expval{\sigma_i \otimes \mathbb{I}}\}$ for $i \in \{x,y,z\}$. As a result, the $W^3$ witnesses are experimentally more efficient to implement than a full tomography, which requires local measurements in nine bases. 

\subsection{Minimization and Performance}
Because entanglement is witnessed if any $\expval{W^3_k}<0$, \textcite{riccardi_optimal_2020} proposed a protocol in which the relevant Stokes parameters are determined from measurements and then the expectation value of each witness is numerically minimized with respect to $\theta$. We find that the entire set of these witnesses may be analytically minimized with respect to $\theta$ (see App.~\ref{sec:W3RepMin} for details). For example,
 \begin{equation} \label{eqn:W3_1_Minimized}
 \begin{aligned}
    \expval{W^3_1}_\text{min} &= \frac{1}{4} \Big[ 1 + S_{zz} \\&- \Big({(S_{xx} + S_{yy})^2+ (S_{z0} + S_{0z})^2}\Big)^{\frac{1}{2}}\Big]. 
 \end{aligned}
\end{equation}
Note that the $\theta$ of Eq.~\ref{eq:w3phistates} that yields minimized expectation values, since it is a function of the measured Stokes parameters, is specific to each state. 

The minimized witness values of Eq.~\ref{eqn:W3_1_Minimized} and App.~\ref{sec:W3RepMin} add insight beyond computational simplification of the entanglement-witnessing protocol. From the analytic expressions in terms of Stokes parameters, it becomes clear that negative expectation values of $W^3$ witness operators have a simple relationship to violation of the PPT criterion for separability of two-qubit states \cite{HORODECKI19961}. $\rho^{\Gamma}$ is positive semi-definite if and only if all its principal minors have non-negative determinants; $\{\expval{W^3_k}\}<0$ signals violation of this condition for 2x2 minor determinants when $\rho$ is written in the bases aligned with each principal axis of the Bloch sphere. For example, when $\rho$ is written in the standard $\ket{\pm z}\otimes\ket{\pm z}$ basis, one 2x2 minor determinant of $\rho^{\Gamma}$ is
\begin{equation}
    \begin{aligned}
        \frac{1}{4}\Bigl[&(1+S_{zz})^2-(S_{0z}+S_{z0})^2\\
    & -(S_{xx}+S_{yy})^2-(S_{xy}-S_{yx})^2\Bigr],
    \end{aligned}\label{eqn:rho14minordet}
\end{equation}
which can never be larger than the simpler quantity
 \begin{equation}
    \frac{1}{4}\Bigl[(1+S_{zz})^2-\Bigl((S_{0z}+S_{z0})^2 +(S_{xx}+S_{yy})^2\Bigr)\Bigr]\label{eqn:rho14minorpart}.
 \end{equation}
By directly comparing Eq.~\ref{eqn:rho14minorpart} with Eq.~\ref{eqn:W3_1_Minimized}, we can see that $\expval{W^3_1}<0$ implies that this minor determinant is negative and thus the PPT criterion is violated.  

To evaluate the performance of the $W^3$ witnesses, we computationally generate a set of $16.8$ million random entangled density matrices, created following the procedure of \textcite{roik_accuracy_2021}. This method consists of randomly drawing a set of eigenvalues for $\rho$, randomizing the basis in which $\rho$ is diagonal, and then filtering the generated states for non-zero entanglement. From this data set we determine the accuracy of $\{W^3\}$, defined as the percentage of all entangled states they successfully detect, to be $33\%$. Given that entanglement can never be witnessed with measurements in a single basis, this performance for three bases leads naturally to the question of how much improvement is possible by adding a limited set of additional measurements. Such considerations motivate our second set of witnesses which build on the measurement results used in $\{W^3\}$.

\section{$\{W^5\}$ Witnesses}\label{sec:W5}

\subsection{Construction}
To improve witnessing performance, we add a basis pair to the measurements performed for $\{W^3\}$ in order to determine $S_{ij}=\expval{\sigma_i \otimes \sigma_j} \textrm{ and } S_{ji}=\expval{\sigma_j \otimes \sigma_i}$ for one pair $i,j \in \{x,y,z\}$ such that $i \neq j$. This results in a set of witnesses which we denote $\{W^5\}$, as they require local measurements in a total of five bases. 

The $W^5$ witnesses can be derived via rotations of the $W^3$ witnesses on one or both particles about a principal axis of the Bloch sphere. Note that rotations of the $W^3$ witnesses about the $z$, $x$, and $y$ axes require additional measurements of $\{S_{xy},S_{yx}\}, \{S_{yz},S_{zy}\}$ and $\{S_{xz},S_{zx}\}$ respectively. Thus, the $W^5$ witnesses can be partitioned according to the geometric rotation performed (\textit{i.e.}, basis pair added) into three subsets or \textit{triplets}, which we denote as $\{W^5\}_n$ for $n \in \{z,x,y\}$.

The $W^5$ witnesses are defined as follows, using two- and three- parameter projectors:
\begin{equation}
    \{W^5\} = \left\{W_k^5=\left(\dyad{\varphi^5_k}\right)^\Gamma \mid k = 1,\dots, 9\right\} \label{eq:W5SetConst}
\end{equation}
where 
\begin{equation}
    \begin{aligned}
        \ket{\varphi_1^5}=& R_z(\alpha)\otimes \mathbb{I}\ket{\varphi_1^3}\\
        \ket{\varphi_2^5} =& R_z(\alpha)\otimes \mathbb{I}\ket{\varphi_2^3}\\
        \ket{\varphi_3^5} =& R_z(\alpha)\otimes R_z(-\beta)\ket{\varphi_3^3}\\
        \ket{\varphi_4^5} =& R_x(\alpha)\otimes \mathbb{I}\ket{\varphi_3^3}\\
        \ket{\varphi_5^5} =& R_x(\alpha)\otimes \mathbb{I}\ket{\varphi_4^3}\\
        \ket{\varphi_6^5} =& R_x(\alpha)\otimes R_x(-\beta)\ket{\varphi_1^3}\\
        \ket{\varphi_7^5} =& R_y(\alpha)\otimes \mathbb{I}\ket{\varphi_5^3}\\
        \ket{\varphi_8^5} =& R_y(\alpha)\otimes \mathbb{I}\ket{\varphi_6^3}\\
        \ket{\varphi_9^5} =& R_y(\alpha)\otimes R_y(\beta)\ket{\varphi_1^3}\\
    \end{aligned}
\end{equation}
for $\theta \in [0,\frac{\pi}{2}]$, $\alpha \in [0,2\pi]$, and $\beta \in [0,2\pi]$. The signs of rotation angles above have been chosen so that the resulting $W^5_k$ operators are expressed in terms of their corresponding $W^3_j$ operators with positive angles. For example,
\begin{equation}
    W^5_6 = \Big(R_x(\alpha)\otimes R_x(\beta)\Big)W_1^3\Big(R_x(\alpha)\otimes R_x(\beta)\Big)^\dagger.
\end{equation}
As with $\{W^3\}$, only parameter values leading to entangled states $\ket{\varphi^5_k}$ yield actual witnesses. Since invalid parameter values give $\Tr\{W^5\rho\}\geq 0$ for all $\rho$, this restriction can be ignored during minimization.

Note the triplet groupings are
\begin{equation}
    \begin{aligned}
        \{W^5\}_z =& \{W_1^5,W_2^5,W_3^5\}\\
        \{W^5\}_x =& \{W_4^5,W_5^5,W_6^5\}\\
        \{W^5\}_y =& \{W_7^5,W_8^5,W_9^5\}.
    \end{aligned}
\end{equation}
\subsection{Minimization and Performance \label{sec:W5Min}}
Six of the nine $W^5$ witnesses may be analytically minimized in the same fashion as $\{W^3\}$. The details and final minimized forms are contained in App.~\ref{sec:W5RepMin}. As an example, the minimized expectation value of $W^5_1$ can be given in terms of Stokes parameters as
\begin{equation}\label{eq:W5_1_minimized}
     \begin{aligned}
     \expval{W_{1}^5}_\text{min} &= \frac{1}{4} \Big[ 1 + S_{zz} -
     \Big( (S_{xx} + S_{yy} ) ^2 \\&+ (S_{z0} + S_{0z})^2 + (S_{xy} - S_{yx})^2\Big)^\frac{1}{2} \Big].
     \end{aligned}
\end{equation}
The analytic minimization for the first two $W^5$ witnesses in each triplet can be found in Eq.~\ref{eq:w5s_minimized}. For the third witnesses in each triplet, which are generated by rotations on both particles, we have found no analytic minimization. We are, however, able to analytically minimize these witness values with respect to $\theta$, resulting in a simplified form with two free parameters seen in Eq.~\ref{eq:W5thirdPmin}. In evaluating the witnesses for specific entangled states, we rely on SciPy minimization to compute the minimized witness expectation values for the third witness in each triplet  \cite{2020SciPy-NMeth}.  

The first two $W^5$ witnesses in each triplet check for negativity of 2x2 minors of $\rho^{\Gamma}$, written in the $z$, $x$, or $y$ bases, via the full criterion of Eq.~\ref{eqn:rho14minordet} rather than the weakened criterion of Eq.~\ref{eqn:rho14minorpart}. The third witness in each triplet checks the weakened criterion, but in $\ket{\pm{n_A}}\otimes\ket{\pm{n'_B}}$ bases where $n_A$ and $n'_B$ lie anywhere in the $xy$ plane (for $W^5_3$), $yz$ plane (for $W^5_6$), or $xz$ plane (for $W^5_9)$.

To analyze the performance of $\{W^5\}$, we use the same computationally generated set of states described in Sec.~\ref{sec:W3}. We find that 79\% of all entangled states in our sample are detected by at least one $W^5$. However, evaluating all three $\{W^5\}_n$
triplets would require all three additional cross-basis measurement pairs, \textit{i.e.}, a full two-qubit tomography. In that case, one could instead reconstruct the density matrix and apply a
complete entanglement criterion. Thus, the 79\% success rate should be interpreted as an oracle benchmark: it represents the
best possible performance of the $W^5$ witnesses if the correct triplet could be selected for each incoming state. The practical goal is therefore to approach this benchmark while measuring only \textit{one} adaptively chosen $\{W^5\}_n$ triplet.

\section{Adaptive Choice Methods}\label{sec:adaptive_choice}
We present two different choice methods that use the measurements already needed for $\{W^3\}$ to predict which $W^5$ triplet is likely to witness entanglement. One method is analytical (the population method) and the other is a five-layer neural network (NN5). The performance of $\{W^5\}$ and the overall protocol depends on the success rate of the adaptive choice method that determines which $\{W^5\}_n$ triplet is used. As such, we analyze the performance of the adaptive protocol while employing these two different choice methods.

\subsection{Population Method}
\begin{table}[b]
    \centering
    \begin{tabular}{|c|c|c|}
    \hline
        Population & $\{\sigma_i, \sigma_j\}$ pair & $\{W^5\}_n$ Triplet\\
    \hline
        $|S_{zz}|$ & $\{\sigma_x, \sigma_y\}$ & $\{W^5\}_z$ \\
    \hline
        $|S_{xx}|$ & $\{\sigma_y, \sigma_z$ \}&  $\{W^5\}_x$ \\
    \hline
        $|S_{yy}|$ & $\{\sigma_x, \sigma_z\}$&$\{W^5\}_y$ \\
    \hline
    \end{tabular}
    \caption{Population method: a $\{W^5\}_n$ triplet is chosen based on which $S_{nn}$ has the maximum magnitude.}
    \label{tab:pop}
\end{table}

The population method is an analytical choice algorithm which chooses a $\{W^5\}_n$ triplet by comparing the relative magnitudes of three Stokes parameters used to find $\{\expval{W^3}\}$: $|S_{zz}|,|S_{xx}|,$ and $|S_{yy}|$. If $S_{nn}$ has the largest magnitude, the triplet $\{W^5\}_n$ is chosen. This logic is mapped in Table~\ref{tab:pop}.

This choice is motivated by the analytic minimization of the first two witnesses in each triplet. As seen in Eq.~\ref{eq:w5s_minimized}, the positive component of the first two minimized witnesses in the $\{W^5\}_n$ triplet are directly dependent on $\pm S_{nn}$. Thus, maximizing $|S_{nn}|$ minimizes the positive component of at least one witness belonging to the $\{W^5\}_n$ triplet, so that a large range of the two cross-basis Stokes parameters to be measured next yield a negative expectation value.

\subsection{Five-Layer Neural Network (NN5)}

Additionally, we present a computational choice model to make the adaptive selection of a $\{W^5\}$ triplet. The model is a five-hidden-layer neural network constructed and trained using Keras \cite{chollet2015keras}. Each dense layer uses a ReLU activation, with a sigmoid activation on the final output, and is trained using the Adam optimizer with binary cross entropy as the loss function. The network takes as input the measurement results in the $xx, yy, \text{and } zz$ bases and predicts which of the three triplet groupings should yield the most negative witness value; the model output is a prediction vector $\va{p}=(p_z,p_x,p_y)$ with exactly one non-zero entry corresponding to the predicted triplet.
 The goal vector $\va{v}=(v_z,v_x,v_y)$ is a three-dimensional multi-hot encoded vector with entry $v_n$ assigned a value of 1 if any member of $\{W^5\}_n$ witnesses the state and 0 if not. 

To train and validate the neural network we generated 4 million random density matrices using the method of \textcite{roik_accuracy_2021} and used only the subset of states which are witnessed by $\{W^5\}$ but unwitnessed by $\{W^3\}$, of which there were 1.208 million.
Training proceeded in two stages: 966,000 states served as the ``training'' set, used to determine the network weights and biases by optimizing the binary cross-entropy loss; the remaining 242,000 served as the ``validation'' set, used to tune the model hyperparameters \footnote{
The NN5 hyperparameters--the number of neurons per dense layer, the batch size, and the learning rate--were tuned to maximize validation accuracy (the fraction of correctly predicted outcomes on the validation set). Hyperparameter optimization was performed via a grid search using Weights\&Biases (L. Biewald, wandb.com), with each layer's neuron count independently sampled uniformly between 1 and 1800, the learning rate sampled uniformly from $10^{-5}$ to $0.5$, and the batch size ranging from 256 to 4480 in increments of 32.
}.

\subsection{Adaptive Choice Performance}
\begin{figure}[t]
    \centering
    \includegraphics[width=0.90\columnwidth]{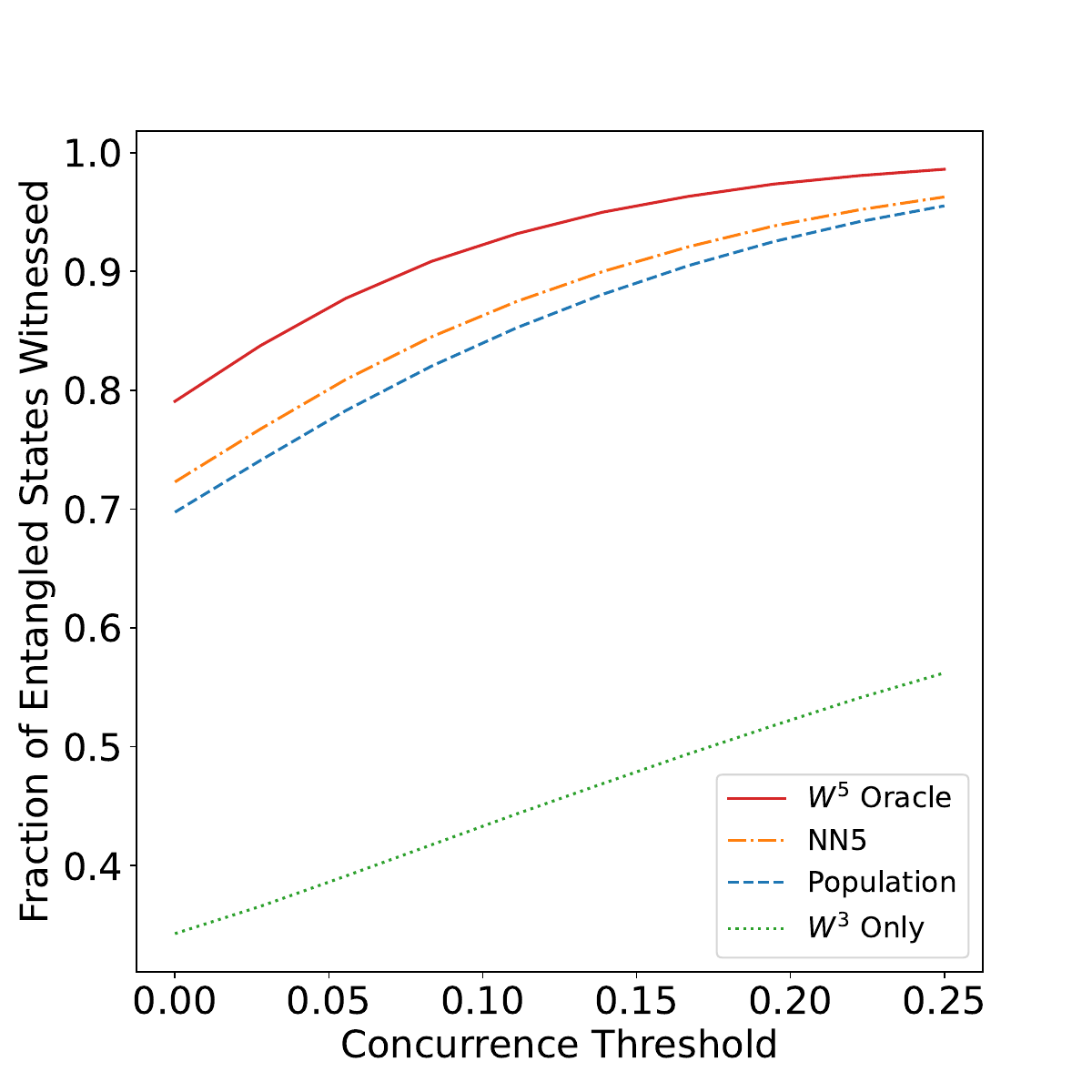}
    \caption{Comparison of the fraction of randomly generated states witnessed by $\{W^3\}$, a $W^5$ oracle, the population method, and the five-layer neural network (NN5). The fraction of successfully witnessed states increases at higher concurrence thresholds. $\{W^3\}$ witnesses $33\%$ of all entangled states and a hypothetical optimal choice of $W^5$ detects $79\%$. The adaptive choice protocol detects $70\%$ via the population method and $72\%$ via NN5.}
    \label{fig:model_perf}
\end{figure}
To evaluate the performance of our overall adaptive protocol when using either the population method or the neural network to make the adaptive choice, we computationally test the model on the same set of generated states from Sec.~\ref{sec:W3}, which are distinct from the states used to train and validate the neural network. The two-step protocol witnesses $72\%$ of all entangled states when using the five-layer neural network and $70\%$ when using the population method. Both success rates fall below the $79\%$ of entangled states witnessed by a $W^5$ oracle, as must be the case since entangled states with identical $\{S_{ii},S_{0i},S_{i0}\}$ are sometimes witnessed by distinct $\{W^5\}_n$. However, both adaptive choice models significantly outperform $\{W^3\}$, and increase the fraction of entangled states witnessed by a very favorable margin relative to the number of additional measurements required.

A protocol may fail to witness very small degrees of entanglement but still be of interest for practical use. Therefore, we also evaluate the adaptive protocol's performance on entangled states above different minimum concurrence thresholds \cite{concurrence1,concurrence2}. As shown in Fig.~\ref{fig:model_perf}, the performance of both choice methods clearly improves when restricted to more strongly entangled states.

\section{Experimental Demonstration\label{sec:expt}}
\begin{figure}[t]
\tikzset{every picture/.style={line width=.75pt}} 
\input{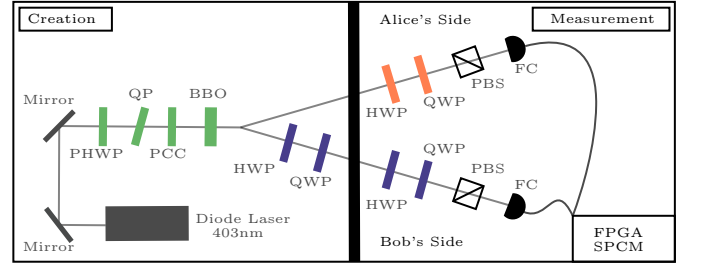}
\caption{Entangled photon pairs are created using $5\SI{0}{\milli\watt}$ $\SI{403}{\nano\meter}$ Diode Laser, PHWP (pump half wave plate), QP (quartz plate), PCC (precompensation crystal), BBO ($\beta$-barium borate crystal), QWP (quarter wave plate), and HWP (half wave plate). The two channels for 806-nm down-converted photons are denoted Alice and Bob; correlated local measurements of specific polarization pairs are achieved with QWP, HWP, PBS (polarizing beam splitter), FC (optical fiber coupler),  SPCM (single-photon counting module) and FPGA for coincidence counting with a 7-ns time window.}
  \label{fig:expt_schematic}
\end{figure}

We demonstrate the success of $\{W^3\}$ and $\{W^5\}$ on specific mixed and pure quantum states composed of polarization-entangled photon pairs produced via type-I spontaneous parametric down conversion using a pair of $\beta$-barium borate crystals (BBO), in an experimental apparatus shown schematically in Fig.~\ref{fig:expt_schematic}. A 403-nm diode laser pump passes through a half wave plate and quartz plate allowing control of the amplitude ratio and phase shift between $\ket{HH}$ and $\ket{VV}$ components of the down-converted state, while the precompensation crystal cancels to first order a BBO-induced phase spread between those components. The remaining quarter and half wave plates on the creation segment of one down-converted channel rotate $\ket{H}$ and $\ket{V}$ polarization states of one photon into an arbitrary elliptical polarization basis. We thus create two-photon entangled states of the general form
\begin{equation}\label{eqn:gen_state}
\begin{aligned}
    \ket{\varphi} = \cos{\frac{\chi}{2}} \ket{H}_A\ket{\alpha}_B + e^{i\gamma}\sin{\frac{\chi}{2}}\ket{V}_A\ket{\alpha_{\perp}}_B
\end{aligned} 
\end{equation}
for any elliptical polarization state $\ket{\alpha}$. The subscripts $A$ and $B$ denote the two members of each entangled pair as photons sent to parties Alice and Bob. To test mixed states, we separately create the pure states in the mixed state's ensemble decomposition and statistically combine measurement results.  

Correlated local measurements on photon pairs are realized with a half and quarter wave plate on each channel to rotate an arbitrary polarization of Alice or Bob's photon into $\ket{H}$, which is then transmitted by a polarizing beam splitter and coupled through multimode fiber to single-photon counting modules. An FPGA-based coincidence counter uses a 7-ns coincidence window to record rates of coincidence counts between the two polarization-selective channels. We use a convention that associates the $z$ axis of the Bloch sphere with polarization states $\ket{H}$ and $\ket{V}$, the $y$ axis with $\ket{R}=\frac{1}{\sqrt{2}}\left(\ket{H}+i\ket{V}\right)$ and $\ket{L}=\frac{1}{\sqrt{2}}\left(\ket{H}-i\ket{V}\right)$, and the $x$ axis with $\ket{D}=\frac{1}{\sqrt{2}}\left(\ket{H}+\ket{V}\right)$ and $\ket{A}=\frac{1}{\sqrt{2}}\left(\ket{H}-\ket{V}\right)$.

\subsection{$\{W^3\}$ Experimental Demonstration}

\begin{figure}[t]
    \centering
    \includegraphics[width=0.90\columnwidth]{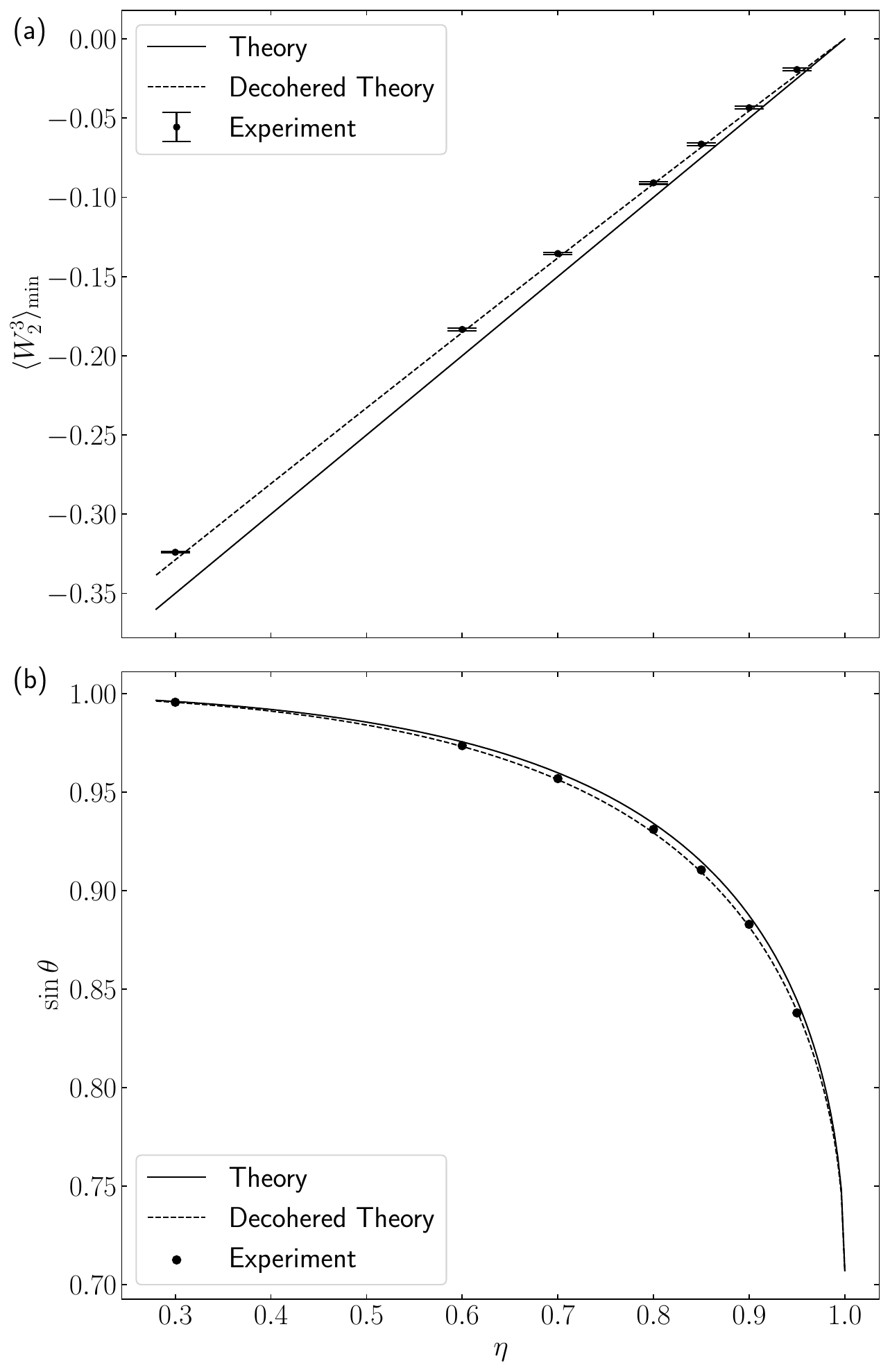}\\
    \caption{Entanglement witnessing via $\expval{W^3_2}$ for amplitude-damped states as a function of damping parameter $\eta$. The decohered theory curve accounts for the experimentally achieved entanglement purity. In (a), $\expval{W^3_2}<0$ shows entanglement is witnessed for all $\eta<1$. The value of $\sin\theta$ that minimizes $\expval{W^3_2}$ varies with $\eta$ as shown in (b). }
        \label{fig:WEx}
\end{figure}

We demonstrate the performance of $\{W^3\}$ on amplitude-damped entangled states of the form
\begin{equation}
    \rho_{ad} = \left(\mathbb{I}\otimes\mathcal{E}\right)\left(\dyad{\Phi^+}\right)
\end{equation}
with $\mathcal{E}(\rho) = A_0\rho A_0^{\dagger}+A_1\rho A_1^{\dagger}$, where $A_0 = \dyad{H} + \sqrt{1-\eta} \dyad{V}$ and $A_1 = \sqrt{\eta}|H\rangle\langle V|$. These states are entangled for any $\eta \in [0,1)$, becoming separable for $\eta=1$. Such states result from realistic noise in many quantum systems (see, \textit{e.g.}, \cite{PhysRevX.13.041022,PhysRevA.107.052409}). \textcite{riccardi_optimal_2020} point out that the optimal value of $\theta$ in Eq.~\ref{eq:w3phistates} for witnessing entanglement in these states varies continuously with the damping parameter $\eta$. As shown in Fig.~\ref{fig:WEx}, $W^3_2$ indeed detects entanglement for any $\eta$;  Fig.~\ref{fig:WEx} also shows how the $\theta$ value minimizing $\expval{W^3_2}$ varies as a function of $\eta$, confirming the observation that nontrivial $\theta$ values are critical for witnessing mixed states of interest.

The primary source of infidelity in our experimental results stems from the presence of residual decoherence between $\ket{HH}$ and $\ket{VV}$ components of our down-converted state. This spread of the phase angle $\gamma$ in Eq.~\ref{eqn:gen_state} limits our experimental maximally entangled states to approximately 94.5\% purity. 
For a target pure state $\ket{\varphi}$ as in Eq.~\ref{eqn:gen_state}, we model this issue with a \textit{decohered theory} state
\begin{equation}
    \begin{aligned}
         \rho_{\text{dec}} = \ &p \ket{\varphi}\bra{\varphi}  \\
     &+ (1-p)\cos^2 \frac{\chi}{2}  \ket{H\alpha}\bra{H\alpha} \\
     &+ (1-p)\sin^2 \frac{\chi}{2}  \ket{V\alpha_\perp}\bra{V\alpha_\perp},
    \end{aligned}
\end{equation}
where $p$ is the experimentally measured purity. In Figs.~\ref{fig:WEx}-\ref{fig:w5tPure}, we plot minimized witness expectation values for our experimental results, target-state theoretical predictions, and these decohered theoretical predictions which account for a dominant source of experimental inaccuracy.

\subsection{$\{W^5\}$ Experimental Demonstration} 
We demonstrate success of the adaptive protocol and $W^5$ witnesses on a set of mixed states given in Eq.~\ref{eqn:w5t1_mixedState}-\ref{eqn:w5t3_mixedState} as well as a set of pure states given in Eq.~\ref{eqn:w5t1_pureState}-\ref{eqn:w5t3_pureState}: 
\begin{equation}
\begin{aligned}
    \ket{\psi_1} &= \cos{(\frac{\chi}{2})}\ket{HH} - i\sin{(\frac{\chi}{2})}\ket{VV} \\
    \ket{\phi_1} &= \cos{(\frac{\chi}{2})}\ket{HV} - i\sin{(\frac{\chi}{2})}\ket{VH} \\
    \rho_{1} &= 0.65\ket{\psi_1}\bra{ \psi_1} + 0.35\ket{\phi_1}\bra{\phi_1}
\end{aligned} \label{eqn:w5t1_mixedState}
\end{equation}
\begin{equation}
\begin{aligned}
    \ket{\psi_{2}} &= \cos{(\frac{\chi}{2})}\ket{HL} - i\sin{(\frac{\chi}{2})}\ket{VR} \\
    \ket{\phi_{2}} &= \cos{(\frac{\chi}{2})}\ket{HA} - i\sin{(\frac{\chi}{2})}\ket{VD} \\
    \rho_{2} &= 0.65\ket{\psi_2}\bra{\psi_2} + 0.35\ket{\phi_2}\bra{\phi_2}
\end{aligned} \label{eqn:w5t2_mixedState}
\end{equation}
\begin{equation}
\begin{aligned}
\ket{\psi_3} &= \cos{(\frac{\chi}{2})}\ket{HL} - \sin{(\frac{\chi}{2})}\ket{VR} \\
 \ket{\phi_3} &= \cos{(\frac{\chi}{2})}\ket{HD} - \sin{(\frac{\chi}{2})}\ket{VA} \\
    \rho_{3} &= 0.65\ket{\psi_3}\bra{\psi_3} + 0.35\ket{\phi_3}\bra{\phi_3}
\end{aligned} 
\label{eqn:w5t3_mixedState}
\end{equation}
\begin{equation}
\begin{aligned}
    \ket{\beta}&= \frac{\sqrt{3}}{2}\ket{H} +\frac{i}{2} \ket{V}\\
    \ket{\beta_\perp} &=\frac{1}{2}\ket{H} - \frac{\sqrt{3}}{2}i\ket{V}\\
    \ket{\psi_4} &= \cos{(\frac{\chi}{2})}\ket{H\beta} -\sin{(\frac{\chi}{2})}\ket{V\beta_\perp}
\end{aligned} \label{eqn:w5t1_pureState}
\end{equation}
\begin{equation}
\begin{aligned}
    \ket{\psi_5} &= \cos{(\frac{\chi}{2})}\ket{HR} + e^{-i\frac{\pi}{6}}\sin{(\frac{\chi}{2})}\ket{VL}
\end{aligned} \label{eqn:w5t2_pureState}
\end{equation}
\begin{equation}
\begin{aligned}
    \ket{\psi_6} &= \cos{(\frac{\chi}{2})}\ket{HA} + e^{-1i}\sin{(\frac{\chi}{2})}\ket{VD}
\end{aligned} \label{eqn:w5t3_pureState}
\end{equation}
\begin{figure*}[t!]
    \centering
    \begin{subfigure}[b]{0.32\textwidth}
        \centering
        \includegraphics[width=1\linewidth]{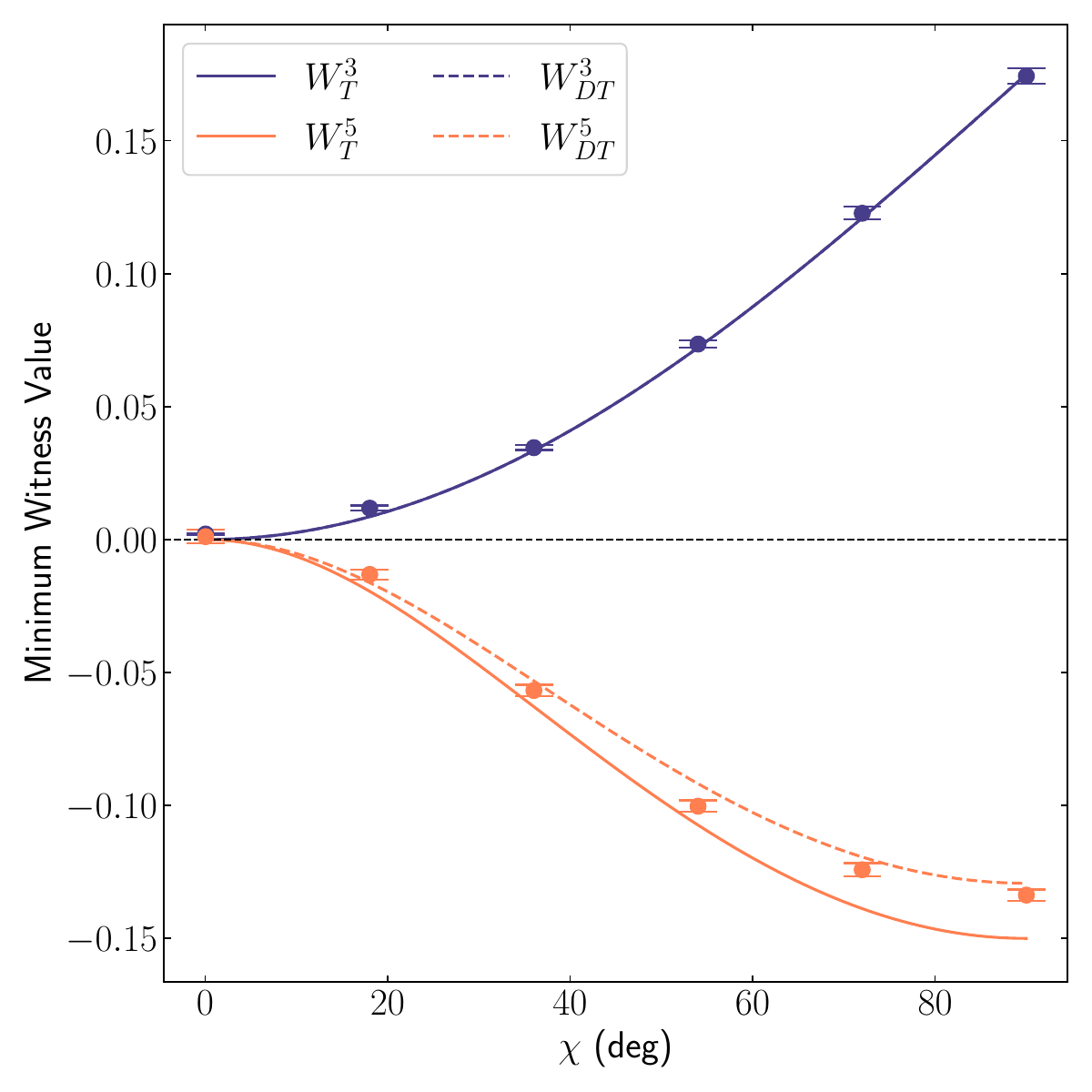}
        \caption{The mixed state of Eq.~\ref{eqn:w5t1_mixedState} witnessed by $\expval{W^5_2}<0$ and $\expval{W^5_3}<0$. \label{fig:w5t1_mixed_plot}}
    \end{subfigure}%
    ~ 
    \begin{subfigure}[b]{0.32\textwidth}
        \centering
        \includegraphics[width=1\linewidth]{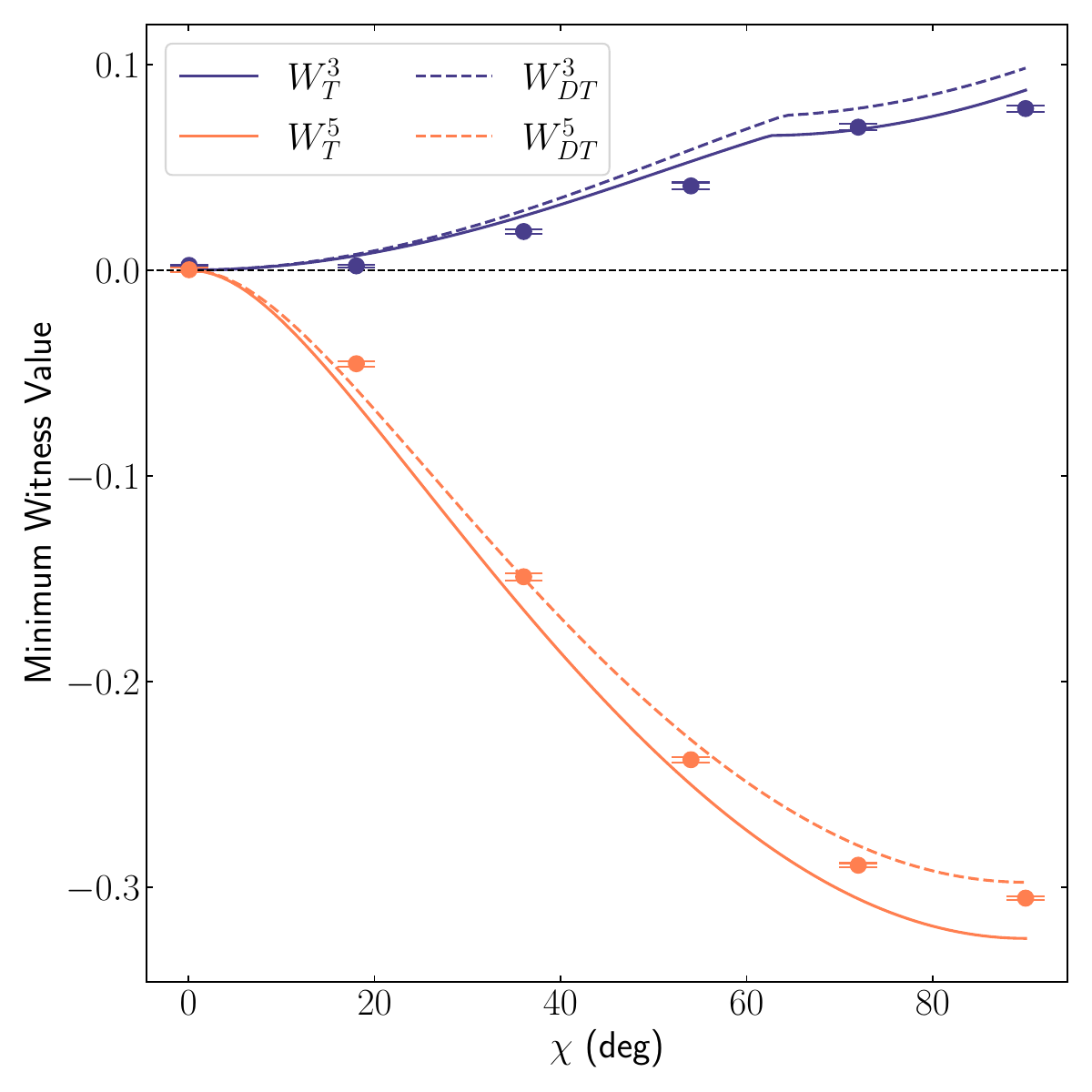}
        \caption{The mixed state of Eq.~\ref{eqn:w5t2_mixedState} witnessed by $\expval{W^5_4}<0$ and $\expval{W^5_6}<0$. \label{fig:w5t2_mixed_plot}}
    \end{subfigure}
     ~ 
    \begin{subfigure}[b]{0.32\textwidth}
        \centering
        \includegraphics[width=1\linewidth]{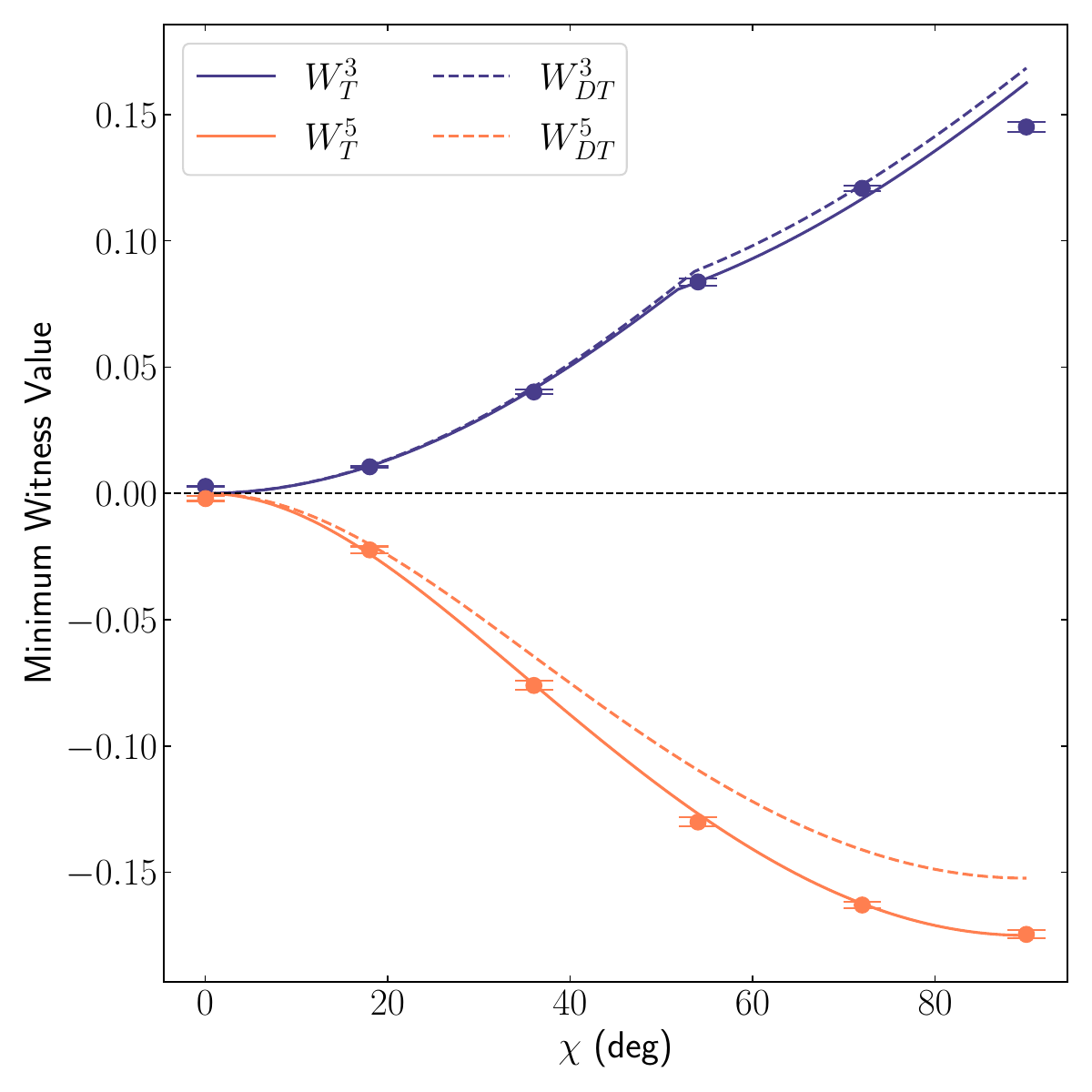}
        \caption{The mixed state of Eq.~\ref{eqn:w5t3_mixedState} witnessed by $\expval{W^5_7}<0$ and $\expval{W^5_9}<0$.\label{fig:w5t3_mixed_plot}}
    \end{subfigure}
    \caption{Minimum $\expval{W^5}$ and $\expval{W^3}$ from experimental data for the mixed states of Eq.~\ref{eqn:w5t1_mixedState}-\ref{eqn:w5t3_mixedState} across a range of $\chi$ values. Each of these mixed states is witnessed by at least one $W^5$ but not by any $W^3$. Curves show theory (T) and decohered theory (DT) predictions for the minimum expectation value in each witness set. This value may not come from the same witness operator at all $\chi$, giving rise to the piecewise appearance of the $\expval{W^3}$ traces in (b)-(c).
    \label{fig:w5tMixed} }
\end{figure*}
\begin{figure*}[t!]
    \centering
    \begin{subfigure}[b]{0.32\textwidth}
        \centering
        \includegraphics[width=1\linewidth]{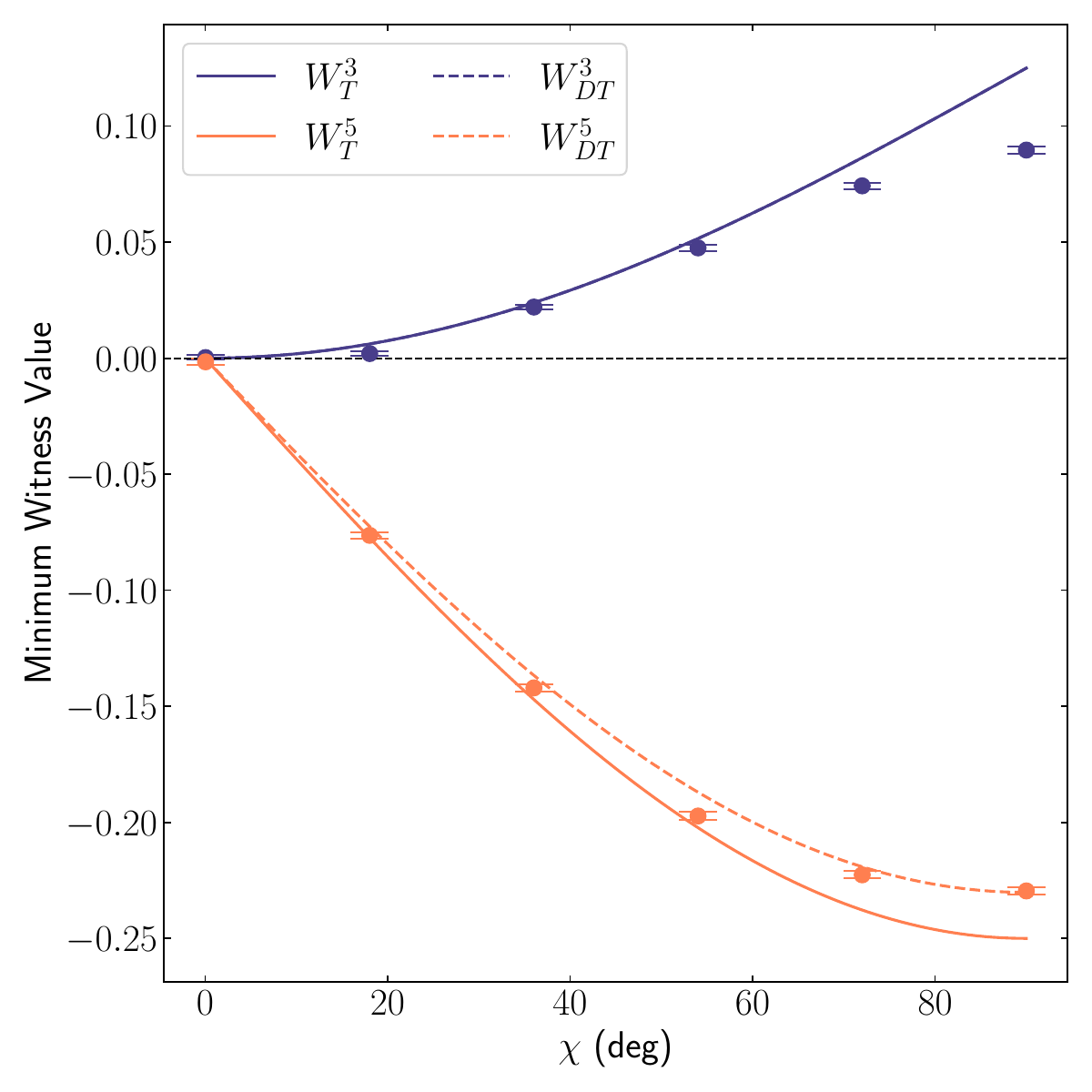}
        \caption{The pure state of Eq.~\ref{eqn:w5t1_pureState} witnessed by $\expval{W^5_2}<0$ and $\expval{W^5_3}<0$.\label{fig:w5t1_pure_plot}}
    \end{subfigure}%
    ~ 
    \begin{subfigure}[b]{0.32\textwidth}
        \centering
        \includegraphics[width=1\linewidth]{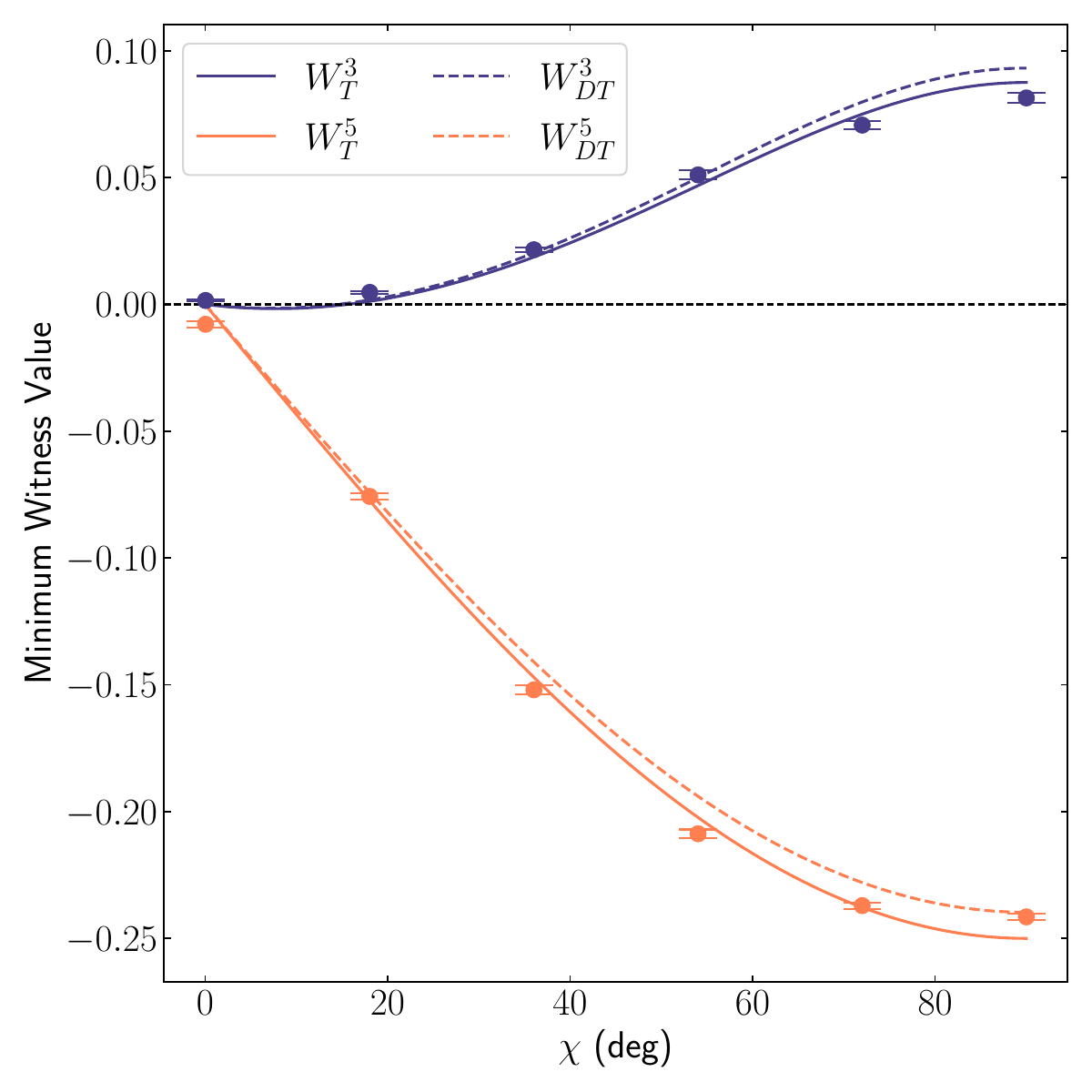}
        \caption{The pure state of Eq.~\ref{eqn:w5t2_pureState} witnessed by $\expval{W^5_4}<0$ and $\expval{W^5_6}<0$. \label{fig:w5t2_pure_plot}}
    \end{subfigure}
     ~ 
    \begin{subfigure}[b]{0.32\textwidth}
        \centering
        \includegraphics[width=1\linewidth]{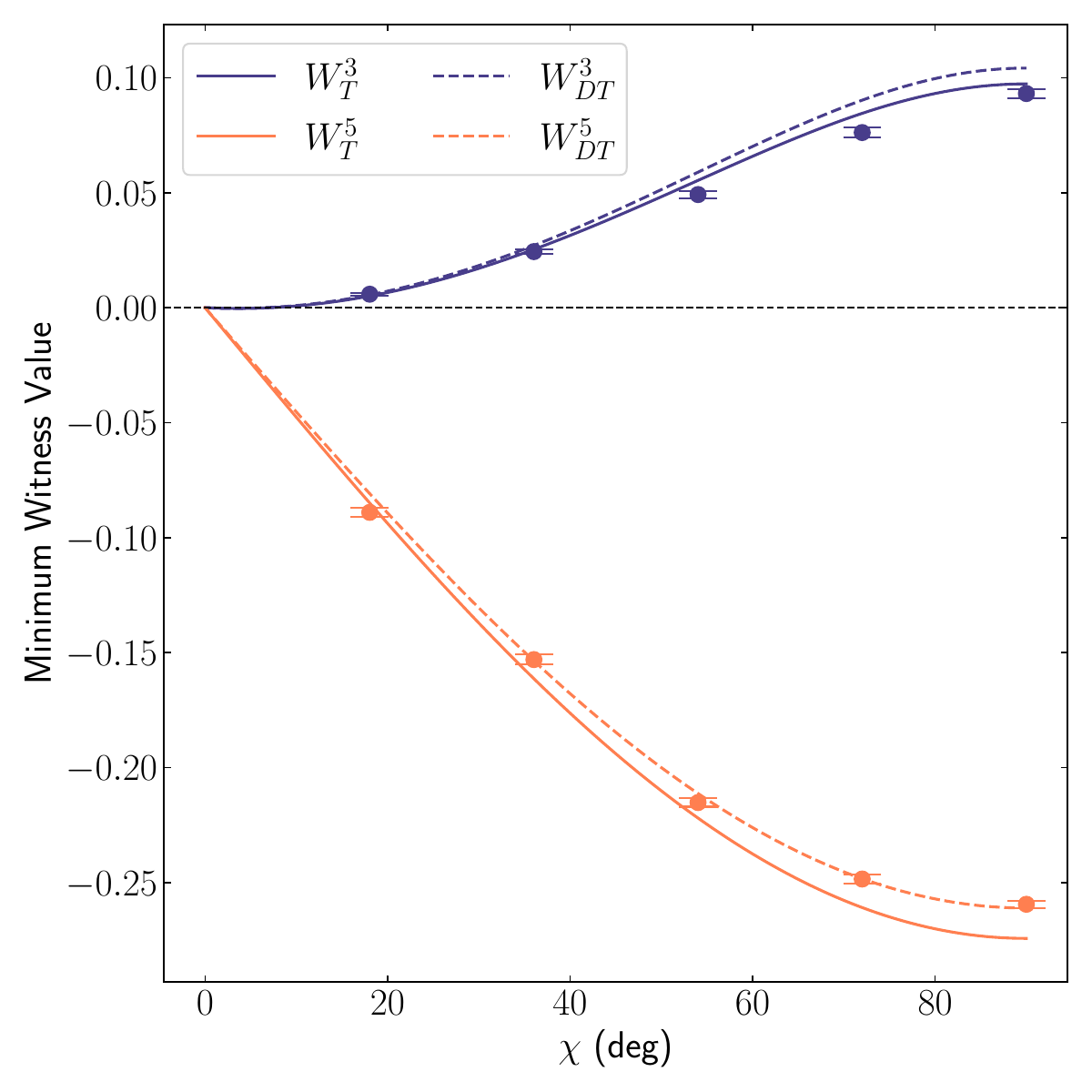}
        \caption{The pure state of Eq.~\ref{eqn:w5t3_pureState} witnessed by $\expval{W^5_7}<0$ and $\expval{W^5_9}<0$. \label{fig:w5t3_pure_plot}}
    \end{subfigure}
    \caption{Minimum $\expval{W^5}$ and $\expval{W^3}$ from experimental data for the pure states of Eq.~\ref{eqn:w5t1_pureState}-\ref{eqn:w5t3_pureState} across a range of $\chi$ values. Each of these pure states is witnessed by at least one $W^5$ but not by any $W^3$. Curves show theory (T) and decohered theory (DT), accounting for small known corrections in experimental state calibration.  
    \label{fig:w5tPure}}
\end{figure*}

These states are not detected by any $W^3$ witness, and each state is witnessed by a distinct $\{W^5\}_n$ triplet. Furthermore, both adaptive choice methods discussed in Sec.~\ref{sec:adaptive_choice} choose the correct $\{W^5\}_n$ triplet for each of these states when the state has non-negligible entanglement, i.e. $\chi\not \approx 0$. These measurements illustrate cases in which the two-step adaptive protocol can be used to efficiently detect entangled states.

Note that in Figs.~\ref{fig:w5tMixed} and~\ref{fig:w5tPure}, many experimental witness values fall closer to theory than to decohered theory. We ascribe this to the fact that minimum witness values are very sensitive to the phase $\gamma$ of Eq.~\ref{eqn:gen_state}. As such, the apparent agreement of some data with theory rather than decohered theory is not a result of underestimating experimental purity, but rather a byproduct of a small shift between targeted and experimentally achieved $\gamma$ values.

Fig.~\ref{fig:w5t1_pure_plot} also shows notable variation of some experimental $\expval{W^3}$ values from both theory and decohered theory predictions. We ascribe this at least in part to slight calibration offsets of the waveplates used to rotate Bob's $\ket{H}$ into $\ket{\alpha}$ of Eq.~\ref{eqn:gen_state}, producing experimental inaccuracies to which the $\expval{W^3}$ are especially sensitive. Experimental state tomography shows that in generating Eq.~\ref{eqn:w5t2_pureState} (Fig.~\ref{fig:w5t2_pure_plot}), rather than $\ket{R}=\cos{\theta} \ket{H} + i\sin{\theta} \ket{V}$ with $\theta=45^{\circ}$, the experimental state had $\theta=42^{\circ}$. For Eq.~\ref{eqn:w5t3_pureState} (Fig.~\ref{fig:w5t3_pure_plot}), rather than $\ket{A}
=\cos{\theta} \ket{H} - \sin{\theta} \ket{V}$ with $\theta=45^{\circ}$, the experimental state had $\theta=44^{\circ}$. The theory and decohered theory curves of Figs.~\ref{fig:w5t2_pure_plot} and \ref{fig:w5t3_pure_plot} account for these state corrections. The calibration procedure for the state of Eq.~\ref{eqn:w5t1_pureState} was notably more complicated than for the other pure states, so we have not corrected the theory curves of Fig.~\ref{fig:w5t1_pure_plot} with any simple model of the experimental offsets. However, calibration offsets of comparable size to those modeled in Figs.~\ref{fig:w5t2_pure_plot} and~\ref{fig:w5t3_pure_plot} can account for the deviations shown in our results. 

Throughout Figs.~\ref{fig:w5tMixed} and~\ref{fig:w5tPure}, the overall success of the adaptive witnessing protocol is clear despite the small experimental inaccuracies we have discussed. Positive $\expval{W^3}$ indicate entanglement is not witnessed with the first three measurement bases while robustly negative $\expval{W^5}$ show clear entanglement witnessing with two additional, adaptively selected, measurement bases.

\section{Conclusions}\label{sec:Concl}

We have presented a two-step method for detecting entanglement of an unknown two-qubit state using a limited set of correlated local measurements. A set of six witnesses previously identified by \textcite{riccardi_optimal_2020}, denoted $\{W^3\}$, requiring local measurements in three bases, can be analytically minimized and are shown to detect $33\%$ of randomly generated entangled states.

We have extended this set with three additional sets of witnesses, denoted $\{W^5\}_n$. Each set consists of three distinct witnesses, which all require correlated local measurements in two additional bases. Our protocol adaptively chooses which set of these witnesses to evaluate using the measurement results already employed to find $\{\expval{W^3}\}$. Overall, the $W^5$ witnesses detect $79\%$ of entangled states, and our two proposed adaptive choice methods detect $70\%$ (analytic choice model) and $72\%$ of entangled states (neural network) while only requiring measurements in five of the nine bases needed for full quantum tomography. We have experimentally demonstrated success cases for our two-step protocol using polarization-entangled photon pairs. 
    
The adaptive choice procedure provides an efficient framework for witnessing two-qubit entanglement. We are currently working to extend the adaptive measurement procedure to a third step, continuing to analyze the fraction of entangled states witnessed relative to the fraction of a full tomography performed. It would also be interesting to explicitly model uncertainty in measurement statistics, and explore tradeoffs in entanglement witnessing efficiency between measurements in new bases and longer measurement times in a smaller set of bases.

\begin{acknowledgments}
The authors thank Laney Goldman, Arianna Meinking, and Richard Zheng for helpful conversations. This work was supported by the Harvey Mudd College (HMC) Physics, Campbell, Class of '69, and Vandiver Summer Research Funds, and the HMC Donnelly Experiential Learning Fund.

The data and code supporting the findings of this article are openly available \cite{dataAndCode}.
\end{acknowledgments}

\appendix

\section{$\{W^3\}$ Alternate Representation and Minimization\label{sec:W3RepMin}}
The $W^3$ witnesses from Eq.~\ref{eq:RicardiW3Set} can be expressed in terms of the canonical Pauli operators as 
\begin{equation} \label{eq:W3Wits}
    \begin{aligned}
    W^3_1 =& \frac{1}{4} [\mathbb{I}\otimes\mathbb{I} + \sigma_z \otimes \sigma_z + \cos(2\theta) (\sigma_x \otimes \sigma_x + \sigma_y \otimes \sigma_y)\\&+ \sin (2\theta) (\sigma_z \otimes\mathbb{I}+\mathbb{I}\otimes \sigma_z)], \\ 
    W^3_2 =& \frac{1}{4} [\mathbb{I}\otimes\mathbb{I} - \sigma_z \otimes \sigma_z + \cos(2\theta) (\sigma_x \otimes \sigma_x - \sigma_y \otimes \sigma_y)\\&+ \sin (2\theta) (\sigma_z \otimes\mathbb{I}-\mathbb{I}\otimes \sigma_z)],\\ 
    W^3_3 =& \frac{1}{4} [\mathbb{I}\otimes\mathbb{I}+ \sigma_x \otimes \sigma_x + \cos(2\theta) (\sigma_z \otimes \sigma_z + \sigma_y \otimes \sigma_y)\\&+ \sin (2\theta) (\sigma_x \otimes\mathbb{I}+ \mathbb{I} \otimes \sigma_x)],\\ 
    W^3_4 =& \frac{1}{4} [\mathbb{I}\otimes\mathbb{I}- \sigma_x \otimes \sigma_x + \cos(2\theta) (\sigma_z \otimes \sigma_z - \sigma_y \otimes \sigma_y)\\&- \sin (2\theta) (\sigma_x \otimes \mathbb{I} - \mathbb{I} \otimes \sigma_x)],\\ 
    W^3_5 =& \frac{1}{4} [\mathbb{I}\otimes\mathbb{I} + \sigma_y \otimes \sigma_y + \cos(2\theta) (\sigma_z \otimes \sigma_z + \sigma_x \otimes \sigma_x)\\&+ \sin (2\theta) (\sigma_y \otimes \mathbb{I} + \mathbb{I} \otimes \sigma_y)], \\ 
    W^3_6 =& \frac{1}{4} [\mathbb{I}\otimes\mathbb{I} - \sigma_y \otimes \sigma_y + \cos(2\theta) (\sigma_z \otimes \sigma_z - \sigma_x \otimes \sigma_x)\\&- \sin (2\theta) (\sigma_y \otimes \mathbb{I} -\mathbb{I} \otimes \sigma_y)]. \\ 
    \end{aligned}
\end{equation}
This allows us to analytically minimize their expectation values with respect to $\theta$ using the trigonometric identity 
 \begin{equation} \label{eq:cos_p_sine_eq_cos}
c \cos \theta \pm d \sin \theta = \sqrt{c^2 + d^2} \cos (\theta \mp \phi),
\end{equation}
where $\sin \phi = \frac{d}{\sqrt{c^2+d^2}}$ and $\cos \phi = \frac{c}{\sqrt{c^2+d^2}}$,
to express the resulting minimum in terms of experimentally measurable Stokes parameters $S_{ij} = \expval{\sigma_i \otimes \sigma_j}, \ i,j \in [0,x,y,z]$.
This results in the expressions
 \begin{equation} \label{eqn:W3Wits_Minimized}
    \begin{aligned}
    \expval{W^3_1}_{\text{min}} =& \frac{1}{4} \left[ 1 + S_{zz} - \sqrt{ (S_{xx} + S_{yy})^2+ (S_{z0} + S_{0z})^2}\right], \\ 
   \expval{ W^3_2 }_{\text{min}} =& \frac{1}{4} \left[ 1 - S_{zz} - \sqrt{( S_{xx} - S_{yy})^2+ (S_{z0} - S_{0z})^2}\right], \\
   \expval{ W^3_3 }_{\text{min}} =& \frac{1}{4} \left[ 1 + S_{xx} - \sqrt{( S_{zz} + S_{yy})^2+ (S_{x0} + S_{0x})^2} \right], \\
    \expval{W^3_4}_{\text{min}} =& \frac{1}{4} \left[ 1 - S_{xx} - \sqrt{( S_{zz} - S_{yy})^2+ (S_{x0} - S_{0x})^2}\right], \\
    \expval{W^3_5 }_{\text{min}} =& \frac{1}{4} \left[ 1 + S_{yy} - \sqrt{( S_{zz} + S_{xx})^2+ (S_{y0} + S_{0y})^2}\right], \\
   \expval{ W^3_6}_{\text{min}} =& \frac{1}{4} \left[ 1 - S_{yy} - \sqrt{( S_{zz} - S_{xx})^2+ (S_{y0} - S_{0y})^2}\right]. \\
    \end{aligned}
\end{equation}

\onecolumngrid
\section{$\{W^5\}$ Alternate Representation and Minimization}\label{sec:W5RepMin}

The $W^5$ witnesses from Eq.~\ref{eq:W5SetConst} can be expressed in terms of the canonical Pauli operators as follows.
The $\{W^5\}_z$ witnesses can be written as
\begin{equation}\label{eq:W5zPauli}
    \begin{aligned}
       W_1^5=& \frac{1}{4} [ \mathbb{I}\otimes\mathbb{I} + \sigma_z \otimes \sigma_z + \cos(2\theta) (\cos(\alpha)(\sigma_x \otimes \sigma_x + \sigma_y \otimes \sigma_y)  \\  &  - \sin(\alpha) (\sigma_x \otimes \sigma_y - \sigma_y \otimes \sigma_x))  +    \sin (2\theta) (\sigma_z \otimes \mathbb{I} + \mathbb{I} \otimes \sigma_z)]\\
       W_2^5 =& \frac{1}{4} [ \mathbb{I}\otimes\mathbb{I} - \sigma_z \otimes \sigma_z + \cos(2\theta) (\cos(\alpha)(\sigma_x \otimes \sigma_x - \sigma_y \otimes \sigma_y) \\  & + \sin(\alpha) (\sigma_x \otimes \sigma_y + \sigma_y \otimes \sigma_x))  +   \sin (2\theta) (\sigma_z \otimes \mathbb{I} - \mathbb{I} \otimes \sigma_z)]\\ 
       W_3^5 =& \frac{1}{4} \Big[ \mathbb{I}\otimes\mathbb{I} + \cos(\alpha) \cos(\beta)\sigma_x \otimes \sigma_x  +\sin(\alpha)\sin(\beta) \sigma_y \otimes \sigma_y    \\ &  + \sin(\alpha)\cos(\beta)  \sigma_y \otimes \sigma_x +\cos(\alpha) \sin(\beta)  \sigma_x \otimes \sigma_y  +\cos(2\theta) \Big(\sigma_z \otimes \sigma_z     + \cos(\alpha) \cos(\beta) \sigma_y \otimes \sigma_y \\&+ \sin(\alpha)\sin(\beta) \sigma_x \otimes \sigma_x -  \sin(\alpha)\cos(\beta) \sigma_x \otimes \sigma_y      - \cos(\alpha) \sin(\beta) \sigma_y \otimes \sigma_x \Big)\\&+ \sin (2\theta) \Big( \cos (\alpha)\sigma_x \otimes \mathbb{I} + \sin(\alpha)\sigma_y \otimes \mathbb{I}     +\cos(\beta)\mathbb{I} \otimes \sigma_x + \sin(\beta)\mathbb{I}\otimes \sigma_y\Big)\Big].
    \end{aligned}
\end{equation}
The $\{W^5\}_x$ witnesses can be written as
\begin{equation}\label{eq:W5xPauli}
    \begin{aligned}
        W_4^5 =& \frac{1}{4} [ \mathbb{I}\otimes\mathbb{I} + \sigma_x \otimes \sigma_x + \cos(2\theta) (\cos(\alpha)(\sigma_z \otimes \sigma_z +  \sigma_y \otimes \sigma_y)  \\  &   - \sin(\alpha) ( \sigma_y \otimes \sigma_z - \sigma_z \otimes \sigma_y))+   \sin (2\theta) ( \sigma_x \otimes \mathbb{I}  + \mathbb{I} \otimes \sigma_x)]\\
        W_5^5 =& \frac{1}{4} [ \mathbb{I}\otimes\mathbb{I} - \sigma_x \otimes \sigma_x + \cos(2\theta) (\cos(\alpha)(\sigma_z \otimes \sigma_z -  \sigma_y \otimes \sigma_y) \\  &   - \sin(\alpha) ( \sigma_y \otimes \sigma_z + \sigma_z \otimes \sigma_y))-  \sin (2\theta) ( \sigma_x \otimes \mathbb{I}  - \mathbb{I} \otimes \sigma_x)]\\
        W_6^5 =& \frac{1}{4} \Big[ \mathbb{I}\otimes\mathbb{I} + \cos(\alpha)\cos(\beta)\sigma_z \otimes \sigma_z + \sin(\alpha) \sin(\beta) \sigma_y \otimes \sigma_y   \\  &   -\sin(\alpha) \cos(\beta)\sigma_y \otimes \sigma_z -  \cos(\alpha)\sin(\beta) \sigma_z \otimes \sigma_y+ \cos(2\theta) \Big(\sigma_x \otimes \sigma_x  +\cos(\alpha) \cos(\beta) \sigma_y \otimes \sigma_y  \\&+\sin(\alpha)\cos(\beta) \sigma_z \otimes \sigma_y +  \cos(\alpha)\sin(\beta) \sigma_y \otimes \sigma_z     + \sin(\alpha) \sin(\beta) \sigma_z \otimes \sigma_z\Big)\\&+  \sin (2\theta) \Big(\cos (\alpha)\sigma_z \otimes \mathbb{I}   -\sin(\alpha)\sigma_y \otimes \mathbb{I} +\cos(\beta)\mathbb{I} \otimes \sigma_z - \sin(\beta)\mathbb{I}\otimes \sigma_y\Big)\Big].
    \end{aligned}
\end{equation}
The $\{W^5\}_y$ witnesses can be written as
\begin{equation}\label{eq:W5yPauli}
    \begin{aligned}
        W_7^5 =&\frac{1}{4} [ \mathbb{I}\otimes\mathbb{I} + \sigma_y \otimes \sigma_y  +\cos(2\theta) (\cos(\alpha) (\sigma_z \otimes \sigma_z + \sigma_x \otimes \sigma_x) \\  &   +\sin(\alpha) (\sigma_x \otimes \sigma_z - \sigma_z \otimes \sigma_x)) +  \sin (2\theta) (\sigma_y \otimes \mathbb{I} +  \mathbb{I} \otimes \sigma_y)]\\
        W_8^5 =& \frac{1}{4} [ \mathbb{I}\otimes\mathbb{I} - \sigma_y \otimes \sigma_y  +\cos(2\theta) (\cos(\alpha) (\sigma_z \otimes \sigma_z - \sigma_x \otimes \sigma_x) \\  &   +\sin(\alpha) (\sigma_x \otimes \sigma_z + \sigma_z \otimes \sigma_x)) -  \sin (2\theta) (\sigma_y \otimes \mathbb{I} -  \mathbb{I} \otimes \sigma_y)]\\
        W_9^5 =& \frac{1}{4} \Big[ \mathbb{I}\otimes\mathbb{I} + \cos(\alpha)\cos(\beta)\sigma_z \otimes \sigma_z + \sin(\alpha) \sin(\beta) \sigma_x \otimes \sigma_x  \\  &   + \sin(\alpha) \cos(\beta)\sigma_x \otimes \sigma_z +  \cos(\alpha)\sin(\beta) \sigma_z \otimes \sigma_x    +\cos(2\theta) \Big(\sigma_y \otimes \sigma_y + \cos(\alpha) \cos(\beta) \sigma_x \otimes \sigma_x  \\&-\sin(\alpha)\cos(\beta) \sigma_z \otimes \sigma_x     - \cos(\alpha)\sin(\beta) \sigma_x \otimes \sigma_z + \sin(\alpha) \sin(\beta) \sigma_z \otimes \sigma_z\Big)\\&+  \sin (2\theta) \Big( \cos (\alpha)\sigma_z \otimes \mathbb{I}  +\sin(\alpha)\sigma_x \otimes \mathbb{I} +\cos(\beta)\mathbb{I} \otimes \sigma_z+\sin(\beta)\mathbb{I}\otimes \sigma_x\Big)\Big].
    \end{aligned}
\end{equation}

The expectations values of the first and second witnesses of each triplet can be analytically minimized with respect to $\theta$ and $\alpha$. As an example, we show the steps in analytical minimization of $\expval{W^5_1}$. Expressing $\expval{W^5_1}$ in terms of Stokes parameters gives
\begin{equation}
    \begin{aligned}
        \min_{\alpha, \theta} \expval{W_{1}^5}=&\frac{1}{4} \Big[ 1 + S_{zz} + \cos(2\theta) \Big(\cos(\alpha)(S_{xx} + S_{yy})    - \sin(\alpha) (S_{xy} - S_{yx})\Big)  +    \sin (2\theta) (S_{z0} + S_{0z})\Big]
    \end{aligned}
\end{equation}
Letting 
\begin{equation}
\begin{aligned}
    a &=   1+S_{zz}&
    b &=  S_{xx}+S_{yy} \\
    c &=  S_{z0} + S_{0z}&
    d &= S_{xy}-S_{yx},
\end{aligned}
\end{equation}
the expression simplifies to
\begin{equation}
    \begin{aligned}
        \min_{\alpha, \theta} \expval{W_{1}^5} =  \min_{\alpha, \theta} & \frac{1}{4} \Big[ a + \cos(2\theta) (b\cos(\alpha)   - d\sin(\alpha) )   +c\sin (2\theta)\Big].
    \end{aligned}
\end{equation}
Using the identity of Eq.~\ref{eq:cos_p_sine_eq_cos} to minimize over $\alpha$, we obtain
\begin{equation}
  \min_{\alpha, \theta} \expval{W_{1}^5} =  \min_{\theta}  \langle\frac{1}{4} [ a + \cos(2\theta) (\sqrt{b^2+d^2}) +  c\sin (2\theta)]\rangle.
\end{equation}
Applying Eq.~\ref{eq:cos_p_sine_eq_cos} again to minimize with respect to $2\theta$, and substituting in the Stokes parameters, we obtain a closed form minimized value for $\expval{W^5_1}$. Using this method, the first two witnesses in each triplet have minimum expectation values, in terms of Stokes parameters, as follows:
\begin{equation}\label{eq:w5s_minimized}
\begin{aligned}
     \expval{W_{1}^5}_\text{min} &= \frac{1}{4} \left[ 1 + S_{zz} - \sqrt{ (S_{xx} + S_{yy} ) ^2 + (S_{z0} + S_{0z})^2 + (S_{xy} - S_{yx})^2} \right] \\
    \expval{W_{2}^5}_\text{min} &= \frac{1}{4} \left[1 - S_{zz}  - \sqrt{(S_{xx} - S_{yy})^2 + (S_{z0} - S_{0z})^2 + (S_{xy}+S_{yx})^2} \right] \\
    \expval{W_{4}^5}_\text{min} &=  \frac{1}{4} \left[1  + S_{xx}  - \sqrt{(S_{zz} + S_{yy})^2 + (S_{0x} + S_{x0})^2 + (S_{yz}-S_{zy})^2} \right] \\
     \expval{W_{5}^5}_\text{min} &= \frac{1}{4} \left[1  - S_{xx}  - \sqrt{(S_{zz} - S_{yy})^2 + (S_{0x} - S_{x0})^2 + (S_{yz}+S_{zy})^2} \right] \\
    \expval{W_{7}^5}_\text{min} &= \frac{1}{4}\left[ 1  + S_{yy} - \sqrt{(S_{zz} + S_{xx} )^2 + (S_{y0} + S_{0y})^2 + (S_{zx} - S_{xz})^2}\right] \\
    \expval{W_{8}^5}_\text{min} &= \frac{1}{4}\left[ 1  - S_{yy} - \sqrt{(S_{zz} - S_{xx} )^2 + (S_{y0} - S_{0y})^2 + (S_{zx} + S_{xz})^2}\right].
\end{aligned}
\end{equation}
The third witness in each triplet cannot be fully analytically minimized in this way; however, each can be minimized with respect to $2\theta$ to yield the following simplified forms:

\begin{equation}
    \begin{aligned} \label{eq:W5thirdPmin}
    \min_{\theta} \expval{W^5_{3}}  =& \frac{1}{4} \Big[ 1 + \cos(\alpha) \cos(\beta)S_{xx}  +\sin(\alpha)\sin(\beta) S_{yy}    + \sin(\alpha)\cos(\beta)  S_{yx} +\cos(\alpha) \sin(\beta)  S_{xy}  \\ 
    &- \big[\big( S_{zz}  + \cos(\alpha) \cos(\beta) S_{yy} + \sin(\alpha)\sin(\beta) S_{xx}    -\sin(\alpha)\cos(\beta) S_{xy} - \cos(\alpha) \sin(\beta) S_{yx}\big)^2 \\ 
    &  + \big( \cos (\alpha)S_{x0} + \sin(\alpha)S_{y0}  +\cos(\beta)S_{0x} + \sin(\beta)S_{0y} \big)^2\big]^{\frac{1}{2}}\Big] \\ 
    \min_\theta \expval{W^5_{6}} =& \frac{1}{4} \Big[ 1 + \cos(\alpha)\cos(\beta)S_{zz} + \sin(\alpha) \sin(\beta) S_{yy}   -\sin(\alpha) \cos(\beta)S_{yz} -  \cos(\alpha)\sin(\beta) S_{zy} \\  &- \big[\big(S_{xx}   +\cos(\alpha) \cos(\beta) S_{yy}  +\sin(\alpha)\cos(\beta) S_{zy}     +\cos(\alpha)\sin(\beta) S_{yz}    + \sin(\alpha) \sin(\beta) S_{zz}\big)^2 \\  &+  \big(\cos (\alpha)S_{z0}-\sin(\alpha)S_{y0} +\cos(\beta)S_{0z} - \sin(\beta)S_{0y}\big)^2\big]^\frac{1}{2} \Big] \\ 
    \min_\theta \expval{W^5_{9}} =&   \frac{1}{4} \Big[ 1 + \cos(\alpha)\cos(\beta)S_{zz} + \sin(\alpha) \sin(\beta) S_{xx}     + \sin(\alpha) \cos(\beta)S_{xy} +  \cos(\alpha)\sin(\beta) S_{zx} \\  &  -\big[(S_{yy} + \cos(\alpha) \cos(\beta) S_{xx}  -\sin(\alpha)\cos(\beta) S_{zx}   - \cos(\alpha)\sin(\beta) S_{xy} + \sin(\alpha) \sin(\beta) S_{zz})^2  \\ &+   ( \cos (\alpha)S_{z0}  +\sin(\alpha)S_{x0} +\cos(\beta)S_{0z}+\sin(\beta)S_{0x})^2\big]^\frac{1}{2}\Big].
\end{aligned}
\end{equation}
\twocolumngrid

\bibliography{references.bib}
\end{document}